\documentclass[english]{article}
\usepackage[T1]{fontenc}
\usepackage[latin9]{inputenc}
\usepackage{amsbsy}
\usepackage{amssymb}
\usepackage{esint}

\makeatletter

\newcommand{\noun}[1]{\textsc{#1}}

\makeatother

\usepackage{babel}
\begin{document}

\title{Probability of Error for Detecting a Change in a Parameter, Total
Variation of the Posterior Distribution, and Bayesian Fisher Information}

\author{Eric Clarkson}
\maketitle
\begin{abstract}
The van Trees inequality relates the Ensemble Mean Squared Error of
an estimator to a Bayesian version of the Fisher Information. The
Ziv-Zakai inequality relates the Ensemble Mean Squared Error of an
estimator to the Minimum Probability of Error for the task of detecting
a change in the parameter. In this work we complete this circle by
deriving an inequality that relates this Minimum Probability of Error
to the Bayesian version of the Fisher Information. We discuss this
result for both scalar and vector parameters. In the process we discover
that an important intermediary in the calculation is the Total Variation
of the posterior probability distribiution function for the parameter
given the data. This total variation is of interest in its own right
since it may be easier to compute than the other figures of merit
discussed here. Examples are provided to show that the inequality
derived here is sharp. 
\end{abstract}

\section{Introduction}

Fisher Information (FI) and the Fisher Information Matrix (FIM) are
fundamental concepts in statistical estimation theory. For a scalar
parameter the well-known Cramer-Rao Bound (CRB) shows that the inverse
of the FI is a lower bound for the variance of an unbiased estimator
of the parameter. For reference we define the FI and state the CRB
for a scalar parameter in Section 2. For a vector parameter the inverse
of the FIM provides a similar lower bound for the covariance matrix
of an unbiased estimator. Less well known is the connection between
FI and the FIM to signal detection theory. For a scalar parameter
we can ask how well we can detect a small change in that parameter
from noisy data associated with it via a conditional Probability Distribution
Function (PDF). The optimal method for detecting such a change is
to compute the likelihood ratio and compare it to a threshold. Such
an observer is called an ideal observer and, by varying the threshold,
we can plot the Receiver Operating Characteristic (ROC) curve for
the ideal observer. The area under this curve, the ideal -observer
AUC, is a figure of merit measuring the quality the data for the task
of detecting the change in the parameter. For small parameter changes,
the ideal-observer AUC is, to first order, proportional to the FI.
This connection between FI and our ability to detect a small change
in a parameter is reviewed briefly in Sections 3. Our main goal in
this work is to find a similar connection between the Bayesian version
of FI and this detection task. For vector parameters the connection
between the ideal-observer AUC for the detection of a small change
in the parameter vector and the FIM is described in Dection 4. We
will also find a similar connection between this detection task and
the Bayesian FIM 

In Section 5 we introduce the Ziv-Zakai inequality in our notation.
The detection task relevant to this inequality is the detection task
we will be considering in the subsequent sections. In this task we
are trying to detect a change in a parameter but we have more information
than we do in the task described in Section 3. In particular, we have
a prior distribution on the parameter and use this to define prior
probabilites for the two parameter values that represent the two hypothese
in the detection task. The ziv-zakai inequality relates the Minimum
Probabiltiy of Error (MPE) for this task to the Ensemble Mean Squared
Error (EMSE) for any estimator of the parameter. The MPE is the probability
of error for the ideal observer in the detection task using a threshold
determined by the prior probabilities of the two hypotheses. In Section
6 we briefly review the van Trees inequality, which relates this EMSE
to the Bayesian FI. These two sections provide context for Section
7, which includes the main result of this paper, an inequality between
the MPE for the Ziv-Zakai detection task and the Baesian FI. In the
process of proving this inequality we introduce an intermediate quantity,
the total variation (TV) of the posterior PDF of the parameter given
the data. This posterior TV may be a useful figure of merit in its
own right since it gives us the first order approximation of the MPE
when the two parameter values are close to each other. The vector-parameter
version of this result is given in Section 8.

When the posterior ODF is unimodal, then the posterior TV is easy
to compute as shown in Section 9. We compute the posterior TV and
Bayesian FI for two examples of unimodal posterior PDFs in Section
10. In Sections 11 and 12 we compute the posterior PDFs, posterior
TVs and Bayesian FIs for two Gaussian examples, one with a scalar
parameter and one with a vector parameter. Finally, in the conclusion
we summarize our results and their implications for the evaluation
of the performance of measurement systems on detection and estimation
tasks.

\section{Fisher Information}

For most of this paper we will be using a scalar parameter $\theta$
and a conditional probability distribution function (PDF) $pr\left(\mathbf{g}|\theta\right)$
for the data vector $\mathbf{g}$. This data vector may, for example,
be the end result of an imaging experiment. However, all of the results
generalize to a vector parameter $\boldsymbol{\theta}$and we will
indicate those generalizations as we proceed. Before getting to the
main new results, and to establish notation, we first review some
concepts relevant to estimation tasks, detection tasks, and the connections
between them. In all that follows angle brackets indicate the probabilistic
expectation and the subscripts on the angle brackets indicate which
random variables or vectors are being avaeraged over and, if needed,
which variables or vectors are held fixed. For example the subscript
$\mathbf{g}|\theta$ means that we are using the conditional PDF $pr\left(\mathbf{g}|\theta\right)$
to average over $\mathbf{g}$with $\theta$ held fixed. 

The Fisher Information (FI) for the parameter of interest $\theta$
is given by the expectation {[}1{]}

\begin{equation}
F\left(\theta\right)=\left\langle \left[\frac{d}{d\theta}\ln pr\left(\mathbf{g}|\theta\right)\right]^{2}\right\rangle _{\mathbf{g}|\theta}.
\end{equation}
If $\hat{\theta}\left(\mathbf{g}\right)$ is an estimator of $\theta$
from the data, then this estimator is unbiased if
\[
\left\langle \hat{\theta}\left(\mathbf{g}\right)\right\rangle _{\mathbf{g}|\theta}=\theta.
\]
The well known Cramer-Rao Bound (CRB) then states that the variance
of any unbiased estimator satisfies {[}1{]}
\begin{equation}
var\left(\hat{\theta}\right)\geq\frac{1}{F\left(\theta\right)}.
\end{equation}
Thus the FI is an important quantity when we are considering estimation
tasks. In the next section we discuss a not-so-well-known relation
between FI and a specific detection task.

\section{FI and ideal-observer AUC}

In this section we introduce a specific binary classification task
that is related to the estimation task in the previous section and
that we will be considering throughout this paper. We suppose that
we are given the data vector $\mathbf{g}$ and told that one of two
hypotheses is true. The hypothesis $H_{1}$ is that $\mathbf{g}$
is a sample drawn from the PDF $pr\left(\mathbf{g}|\theta\right)$,
which we write as $\mathbf{g}\sim pr\left(\mathbf{g}|\theta\right)$.
The hypothesis $H_{1}$ is that $\mathbf{g}\sim pr\left(\mathbf{g}|\tilde{\theta}\right)$.
Then job of the observer is to determine which hypothesis is true.
The optimal observer for this task by many metrics, some of which
we will be discussing below, is the Bayesian ideal observer, also
known simply as the ideal observer. This observer computes the likelihood
ratio {[}2,3{]}

\begin{equation}
\Lambda\left(\mathbf{g}|\theta,\tilde{\theta}\right)=\frac{pr\left(\mathbf{g}|\tilde{\theta}\right)}{pr\left(\mathbf{g}|\theta\right)}
\end{equation}
and compares the result to a threshold that we will call $y$. If
$\Lambda\left(\mathbf{g}|\theta,\tilde{\theta}\right)>y$ then the
ideal observer concludes that$\mathbf{g}\sim pr\left(\mathbf{g}|\tilde{\theta}\right)$,
i.e. that hypothesis $H_{1}$ is true. Otherwise this observer declares
that $\mathbf{g}\sim pr\left(\mathbf{g}|\theta\right)$ and hypothesis
$H_{0}$ is true. 

Due to noise in the data vector quantified by the PDFs $pr\left(\mathbf{g}|\theta\right)$
and $pr\left(\mathbf{g}|\tilde{\theta}\right)$ the ideal observer,
although optimal, is not always right. One possible error is a False
Positive (FP) where $\mathbf{g}\sim pr\left(\mathbf{g}|\theta\right)$
but $\Lambda\left(\mathbf{g}|\theta,\tilde{\theta}\right)>y$. To
find an expression for the probability of an FP outcome, also known
as the False Positive Fraction (FPF) we first note that when $\mathbf{g}\sim pr\left(\mathbf{g}|\theta\right)$
the likelihood ratio $\Lambda=\Lambda\left(\mathbf{g}|\theta,\tilde{\theta}\right)$
is a random variable with a PDF that we will denote by $pr_{0}\left(\Lambda|\theta,\tilde{\theta}\right)$.
The subscript indicates the hypothesis that is in force, and the $\theta$
and $\tilde{\theta}$ after the vertical bar are there because the
function $\Lambda\left(\mathbf{g}|\theta,\tilde{\theta}\right)$,
and hence the PDF for $\Lambda$, depends on both of these variables.
The FPF can now be written as
\begin{equation}
FPF\left(y|\theta,\tilde{\theta}\right)=\int_{y}^{\infty}pr_{0}\left(\Lambda|\theta,\tilde{\theta}\right)d\Lambda.
\end{equation}
The other possible error is a False Negative (FN) where $\mathbf{g}\sim pr\left(\mathbf{g}|\tilde{\theta}\right)$
but $\Lambda\left(\mathbf{g}|\theta,\tilde{\theta}\right)\leq y$.
Using the notation we just described for the FPF, he probability of
an FN outcome, the False Negative Fraction (FNF), is given by
\begin{equation}
FNF\left(y|\theta,\tilde{\theta}\right)=\int_{0}^{y}pr_{1}\left(\Lambda|\theta,\tilde{\theta}\right)dt.
\end{equation}
The True Positive Fraction (TPF) is defined by $TPF\left(y|\theta,\tilde{\theta}\right)=1-FNF\left(y|\theta,\tilde{\theta}\right)$
and is the probabiity that $\Lambda\left(\mathbf{g}|\theta,\tilde{\theta}\right)>y$
when $H_{1}$ is valid. For a given pair $\left(\theta,\tilde{\theta}\right)$
the corresponding Receiver Operating Characteristic (ROC) curve is
a plot of $TPF\left(y|\theta,\tilde{\theta}\right)$ versus $FPF\left(y|\theta,\tilde{\theta}\right)$.
The area under this curve is the ideal-observer AUC, and is a can
be used as a figure of merit for the quality of the data with respect
to the classification task. One advantage of the AUC as a figure of
merit is that it can be estimated from a Two Alternative Forced Choice
(2AFC) test without actually plotting the ROC curve. We will use the
notation $AUC\left(\theta,\tilde{\theta}\right)$ for this area. The
ideal observer detectability $d\left(\theta,\tilde{\theta}\right)$
is an alternative to $AUC\left(\theta,\tilde{\theta}\right)$ as a
figure of merit and is defined by

\begin{equation}
AUC\left(\theta,\tilde{\theta}\right)=\frac{1}{2}+\frac{1}{2}\mathrm{erf}\left[\frac{1}{2}d\left(\theta,\tilde{\theta}\right)\right].
\end{equation}
The ideal-observer AUC always satisfies $0.5\leq AUC\left(\theta,\tilde{\theta}\right)\leq1$
and the ideal-observer detectability is always a non-negative real,
number.

In previous work it was shown that the ideal-observer detectability
is related to the FI in a Taylor series expansion as {[}4,5,6{]}
\begin{equation}
d^{2}\left(\theta,\theta+\triangle\theta\right)=F\left(\theta\right)\left(\triangle\theta\right)^{2}+\ldots
\end{equation}
Thus the FI has an interpretation in terms of detcting a small change
in a parameter. The next term in the Taylor series, the cubic term,
is known and involves the derivative of the FI. Thus it is possible
to estimate the error in the second order approximation to the detectability.
In order to compare this result to some results derived below we want
to relate the FI directly to $AUC\left(\theta,\theta+\triangle\theta\right)$.
To do this we start with the Taylor series for the error function:
\begin{equation}
\mathrm{erf}\left(z\right)=\frac{2}{\sqrt{\pi}}\left(z-\frac{z^{3}}{3}+\ldots\right).
\end{equation}
Using this series we find that
\begin{equation}
AUC\left(\theta,\theta+\triangle\theta\right)=\frac{1}{2}+\frac{1}{2\sqrt{\pi}}\left|\triangle\theta\right|\sqrt{F\left(\theta\right)}+\ldots
\end{equation}
Note that the absolute value here indicates that $AUC\left(\theta,\tilde{\theta}\right)$
is not differentiable with respect to $\tilde{\theta}$ at $\tilde{\theta}=\theta$.
We can however formulate this equation in terms of one-sided derivatives
as
\begin{equation}
\left.\frac{d}{d\tilde{\theta}}AUC\left(\theta,\tilde{\theta}\right)\right|_{\tilde{\theta}=\theta\pm}=\pm\frac{1}{2\sqrt{\pi}}\sqrt{F\left(\theta\right)}
\end{equation}
This formulation can be compared to results below relating the minimum
probability of error on this classification task to the Bayesian version
of the FI.

\section{Vector Version of FI and AUC}

For a $p$-dimensional vector parameter $\boldsymbol{\theta}$ the
$p\times p$ Fisher Information Matrix (FIM) is defined by {[}1{]}
\begin{equation}
\mathbf{F}\left(\boldsymbol{\theta}\right)=\left\langle \left[\nabla_{\boldsymbol{\theta}}\ln pr\left(\mathbf{g}|\boldsymbol{\theta}\right)\right]\left[\nabla_{\boldsymbol{\theta}}\ln pr\left(\mathbf{g}|\boldsymbol{\theta}\right)\right]^{\dagger}\right\rangle _{\mathbf{g}|\boldsymbol{\theta}}.
\end{equation}
This matrix is related to the ideal-observer detectability of a small
change in the parameter vector by the Taylor series expansion {[}4,5,6{]}

\begin{equation}
d^{2}\left(\boldsymbol{\theta},\boldsymbol{\theta}+\triangle\boldsymbol{\theta}\right)=\triangle\boldsymbol{\theta}^{\dagger}\mathbf{F}\left(\boldsymbol{\theta}\right)\triangle\boldsymbol{\theta}+\ldots
\end{equation}
If $\mathbf{u}$is an arbitrary unit vector in parameter space then
we have the one sided derivatives
\begin{equation}
\left.\frac{d}{dt}AUC\left(\boldsymbol{\theta},\boldsymbol{\theta}+t\mathbf{u}\right)\right|_{t=0+}=\frac{1}{2\sqrt{\pi}}\sqrt{\mathbf{u}^{\dagger}\mathbf{F}\left(\boldsymbol{\theta}\right)\mathbf{u}}
\end{equation}
and
\begin{equation}
\left.\frac{d}{dt}AUC\left(\boldsymbol{\theta},\boldsymbol{\theta}+t\mathbf{u}\right)\right|_{t=0-}=-\frac{1}{2\sqrt{\pi}}\sqrt{\mathbf{u}^{\dagger}\mathbf{F}\left(\boldsymbol{\theta}\right)\mathbf{u}}.
\end{equation}
For example, when $p=2$ we can plot $AUC\left(\boldsymbol{\theta},\tilde{\boldsymbol{\theta}}\right)$
as a function of $\tilde{\boldsymbol{\theta}}$ for a fixed $\boldsymbol{\theta}$.
This surface will descend to a point at $\tilde{\boldsymbol{\theta}}=\boldsymbol{\theta}$.
The slope as we descend to or ascend from this singularity in the
direction $\mathbf{u}$ is determined by the quantity $\mathbf{u}^{\dagger}\mathbf{F}\left(\boldsymbol{\theta}\right)\mathbf{u}$.
A larger value for this slope implies that it will be easier to detect
a small change in the parameter vector in that direction from the
data that we have to work with.

\section{The Ziv-Zakai Inequality}

The setting for the results we will be describing below relating the
Minimum Probability of Error (MPE) is the calssification task described
above to the Bayesian FI is the same as the setting for the Ziv-Zakai
inequality that relates the Ensemble Mean Squared error (EMSE) of
an estimator to this same MPE. We will briefly discuss this inequality
in order to introduce this setting and some notation we will be using.
If we have a prior PDF $pr\left(\theta\right)$ on the parameter of
interest then we may define probabi;ites for the two hypotheses $H_{0}$
and $H_{1}$ via

\begin{equation}
Pr_{0}\left(\theta,\tilde{\theta}\right)=\frac{pr\left(\theta\right)}{pr\left(\theta\right)+pr\left(\tilde{\theta}\right)}
\end{equation}
and
\begin{equation}
Pr_{1}\left(\theta,\tilde{\theta}\right)=\frac{pr\left(\tilde{\theta}\right)}{pr\left(\theta\right)+pr\left(\tilde{\theta}\right)}.
\end{equation}
The probability of error for the ideal observer when the threshold
is $y$ is then given by
\begin{equation}
Pr_{0}\left(\theta,\tilde{\theta}\right)\int_{y}^{\infty}pr_{0}\left(\Lambda|\theta,\tilde{\theta}\right)dt+Pr_{1}\left(\theta,\tilde{\theta}\right)\int_{0}^{y}pr_{1}\left(\Lambda|\theta,\tilde{\theta}\right)dt.
\end{equation}
The two terms here correspond to the FP and FN cases. To minimize
the probability of error the optimal threshold is given by 
\begin{equation}
y=y\left(\theta,\tilde{\theta}\right)=\frac{Pr_{0}\left(\theta,\tilde{\theta}\right)}{Pr_{1}\left(\theta,\tilde{\theta}\right)}=\frac{pr\left(\theta\right)}{pr\left(\tilde{\theta}\right)}.
\end{equation}
The MPE in this setting is therefore
\begin{equation}
P_{e}\left(\theta,\tilde{\theta}\right)=Pr_{0}\left(\theta,\tilde{\theta}\right)\int_{y\left(\theta,\tilde{\theta}\right)}^{\infty}pr_{0}\left(\Lambda|\theta,\tilde{\theta}\right)dt+Pr_{1}\left(\theta,\tilde{\theta}\right)\int_{0}^{y\left(\theta,\tilde{\theta}\right)}pr_{1}\left(\Lambda|\theta,\tilde{\theta}\right)dt.
\end{equation}
We always have $0\leq P_{e}\left(\theta,\tilde{\theta}\right)\leq\min\left\{ Pr_{0}\left(\theta,\tilde{\theta}\right),Pr_{1}\left(\theta,\tilde{\theta}\right)\right\} \leq0.5$
since an observer could just decide $H_{0}$ or $H_{1}$ is true every
time. 

The EMSE for an estimator $\hat{\theta}\left(\mathbf{g}\right)$ is
given by
\begin{equation}
EMSE\left(\hat{\theta}\right)=\left\langle \left\langle \left[\hat{\theta}\left(\mathbf{g}\right)-\theta\right]^{2}\right\rangle _{\mathbf{g}|\theta}\right\rangle _{\theta}.
\end{equation}
The usual formulation of the Ziv-Zakai inequality can now be written
as {[}7,8{]}

\begin{equation}
EMSE\left(\hat{\theta}\right)\geq\frac{1}{2}\int_{0}^{\infty}\int_{-\infty}^{\infty}\left[pr\left(\theta\right)+pr\left(\theta+x\right)\right]P_{e}\left(\theta,\theta+x\right)d\theta xdx.
\end{equation}
We have shown elsewhere that by using straightforward changes of variable
and the symmetry of the function $P_{e}\left(\theta,\tilde{\theta}\right)$
this inequality can also be written as
\begin{equation}
EMSE\left(\hat{\theta}\right)\geq\frac{1}{2}\left\langle \int_{-\infty}^{\infty}P_{e}\left(\theta,\tilde{\theta}\right)\left|\tilde{\theta}-\theta\right|d\tilde{\theta}\right\rangle _{\theta}.
\end{equation}
For the curious the derivation of this version of the Ziv-Zakai inequality
is shown in the Appendix. A large value of $P_{e}\left(\theta,\tilde{\theta}\right)$
when $\left|\tilde{\theta}-\theta\right|$ is small is intuitively
expected, as is a small value of $P_{e}\left(\theta,\tilde{\theta}\right)$
when $\left|\tilde{\theta}-\theta\right|$ is large. The Ziv-Zakai
inequality shows that a large value of $P_{e}\left(\theta,\tilde{\theta}\right)$
when $\left|\tilde{\theta}-\theta\right|$ is also large is very bad
for the estimation problem as it will force a large EMSE for any estimator.
We will be showing below that the behavior of $P_{e}\left(\theta,\tilde{\theta}\right)$
when $\left|\tilde{\theta}-\theta\right|$ is small is related to
the total variation of the posterior PDF for $\theta$, and to the
Bayesian FI. 

\section{Bayesian FI and EMSE}

What we have been referring to as the Bayesian FI is given by

\begin{equation}
F=\left\langle F\left(\theta\right)\right\rangle _{\theta}+\left\langle \left[\frac{d}{d\theta}\ln pr\left(\theta\right)\right]^{2}\right\rangle _{\theta}.
\end{equation}
The posterior PDF $pr\left(\theta|\mathbf{g}\right)$ for $\theta$
is defined by the equation $pr\left(\theta|\mathbf{g}\right)pr\left(\mathbf{g}\right)=pr\left(\mathbf{g}|\theta\right)pr\left(\theta\right)$,
where 
\[
pr\left(\mathbf{g}\right)=\int_{-\infty}^{\infty}pr\left(\mathbf{g}|\theta\right)pr\left(\theta\right)d\theta.
\]
In terms of this posterior PDF the Bayesian FI can also be written
as
\begin{equation}
F=\left\langle \left\langle \left[\frac{d}{d\theta}\ln pr\left(\theta|\mathbf{g}\right)\right]^{2}\right\rangle _{\mathbf{g}|\theta}\right\rangle _{\theta}.
\end{equation}
This expression will be useful when we relate $P_{e}\left(\theta,\tilde{\theta}\right)$
to the Bayesian FI. The usual application of the Bayesian FI is the
van Trees inequality, also called the Bayesian CRB, which states that
{[}9,10{]}
\begin{equation}
EMSE\left(\hat{\theta}\right)\geq F^{-1}.
\end{equation}
There are versions of the Ziv-Zakiai inequality and the van Trees
inequality for vector parameters but we will not be discussing those
here. We would be completing the circle started by these two inequalities
if we had a relation between $P_{e}\left(\theta,\tilde{\theta}\right)$
and $F$. This is the subject of the next section and we will find
that an intermediary in this relation is the total variation of the
posterior PDF $pr\left(\theta|\mathbf{g}\right)$.

\section{Bayesian FI and MPE}

The ideal observer for the classification task in the Ziv-Zakai inequality
can be formulated by defining a test statistic $t\left(\mathbf{g}|\theta,\tilde{\theta}\right)$
by

\begin{equation}
t\left(\mathbf{g}|\theta,\tilde{\theta}\right)=\frac{pr\left(\tilde{\theta}|\mathbf{g}\right)}{pr\left(\theta|\mathbf{g}\right)}=\frac{pr\left(\mathbf{g}|\tilde{\theta}\right)}{pr\left(\mathbf{g}|\theta\right)}\frac{pr\left(\tilde{\theta}\right)}{pr\left(\theta\right)}=\frac{pr\left(\tilde{\theta}\right)}{pr\left(\theta\right)}\Lambda\left(\mathbf{g}|\theta,\tilde{\theta}\right)
\end{equation}
and declaring the hypothesis $H_{1}$ to be correct if $t\left(\mathbf{g}|\theta,\tilde{\theta}\right)>1.$
This is obviously equalivalent to using the likelihood ratio test
statistic and the threshold $y\left(\theta,\tilde{\theta}\right)$
given above. Thus this classification scheme acheives the minimum
possible value of the probability of error for this task. We will
use the following notation for certain derivatives. For the test statistic
we have
\begin{equation}
t'\left(\mathbf{g}|\theta\right)=\left.\frac{d}{d\tilde{\theta}}t\left(\mathbf{g}|\theta,\tilde{\theta}\right)\right|_{\tilde{\theta}=\theta}.
\end{equation}
For the conditional and prior PDFs we use
\begin{equation}
pr'\left(\mathbf{g}|\theta\right)=\left.\frac{d}{d\tilde{\theta}}pr\left(\mathbf{g}|\tilde{\theta}\right)\right|_{\tilde{\theta}=\theta}
\end{equation}
and
\begin{equation}
pr'\left(\theta\right)=\left.\frac{d}{d\tilde{\theta}}pr\left(\tilde{\theta}\right)\right|_{\tilde{\theta}=\theta}
\end{equation}
These derivatives are related by the equation
\begin{equation}
t'\left(\mathbf{g}|\theta\right)=\frac{pr'\left(\mathbf{g}|\theta\right)}{pr\left(\mathbf{g}|\theta\right)}+\frac{pr'\left(\theta\right)}{pr\left(\theta\right)}.
\end{equation}
This notation will make the derivation of the main results easier
to follow.

The Bayesian FI can now be written as
\begin{equation}
F=\left\langle \left\langle \left[t'\left(\mathbf{g}|\theta\right)\right]^{2}\right\rangle _{\mathbf{g}|\theta}\right\rangle _{\theta}.
\end{equation}
The function $t'\left(\mathbf{g}|\theta\right)$ can be viewed as
a random variable $t'$ since $\mathbf{g}$ is a random vector with
conditional PDF $pr\left(\mathbf{g}|\theta\right)$ and $\theta$
is a random variable with PDF $pr\left(\theta\right)$. The mean of
this random variable is given by
\begin{equation}
\left\langle \left\langle t'\left(\mathbf{g}|\theta\right)\right\rangle _{\mathbf{g}|\theta}\right\rangle _{\theta}=\left\langle \left\langle \frac{pr'\left(\mathbf{g}|\theta\right)}{pr\left(\mathbf{g}|\theta\right)}\right\rangle _{\mathbf{g}|\theta}+\frac{pr'\left(\theta\right)}{pr\left(\theta\right)}\right\rangle _{\theta}=\left\langle \frac{pr'\left(\theta\right)}{pr\left(\theta\right)}\right\rangle _{\theta}=0
\end{equation}
and we therefore we have $F=var\left(t'\right)$. For fixed $\left(\theta,\tilde{\theta}\right)$
the function $t\left(\mathbf{g}|\theta,\tilde{\theta}\right)$ can
be viewed as a random variable $t$ with conditional PDFs $pr_{0}\left(t|\theta,\tilde{\theta}\right)$
and $pr_{1}\left(t|\theta,\tilde{\theta}\right)$ under the two hypotheses
$H_{0}$ and $H_{1}$ respectively. In terms of these PDFs the MPE
function $P_{e}\left(\theta,\tilde{\theta}\right)$ can be written
as
\begin{equation}
P_{e}\left(\theta,\tilde{\theta}\right)=Pr_{0}\left(\theta,\tilde{\theta}\right)\int_{1}^{\infty}pr_{0}\left(t|\theta,\tilde{\theta}\right)dt+Pr_{1}\left(\theta,\tilde{\theta}\right)\int_{0}^{1}pr_{1}\left(t|\theta,\tilde{\theta}\right)dt.
\end{equation}
We want to compute the derivative of this function with respect to
$\tilde{\theta}$ evaluated at $\tilde{\theta}=\theta$. The magnitude
of this derivative telss us how rapidly the MPE changes as $\tilde{\theta}$
moves away from $\theta$. This in turn tells us how useful the data
is for the Ziv-Zakai classification task when $\tilde{\theta}$ is
close to $\theta$. 

Before we try to compute the derivative in question we need to explain
the relation between the PDFs $pr_{0}\left(t|\theta,\tilde{\theta}\right)$
and $pr_{1}\left(t|\theta,\tilde{\theta}\right)$. The corresponding
PDfs for the likelihood ratio satisfy the relation $pr_{1}\left(\Lambda|\theta,\tilde{\theta}\right)=\Lambda pr_{0}\left(\Lambda|\theta,\tilde{\theta}\right)$.
To see how this property translates to the PDFs $pr_{0}\left(t|\theta,\tilde{\theta}\right)$
and $pr_{1}\left(t|\theta,\tilde{\theta}\right)$ we consider two
random variables $w$ and $x$ that are related by$w=cx$ for some
constant $c$. Then we have the standard relation$pr_{w}\left(w\right)=\left(1/c\right)pr_{x}\left(w/c\right)$.
Now suppose the we have a different PDF $\tilde{pr}_{x}\left(x\right)$
given by$\tilde{pr}_{x}\left(x\right)=xpr_{x}\left(x\right)$. The
corresponding PDF for $w$ is then given by
\begin{equation}
\tilde{pr}_{w}\left(w\right)=\frac{1}{c}\tilde{pr}_{x}\left(\frac{w}{c}\right)=\frac{w}{c^{2}}pr_{x}\left(\frac{w}{c}\right)=\frac{w}{c}pr_{w}\left(w\right).
\end{equation}
Translating this result to the random variable $t$ we have
\begin{equation}
pr_{1}\left(t|\theta,\tilde{\theta}\right)=\frac{pr\left(\theta\right)}{pr\left(\tilde{\theta}\right)}tpr_{0}\left(t|\theta,\tilde{\theta}\right).
\end{equation}
Using this esult we may wrtie the MPE function as
\begin{equation}
P_{e}\left(\theta,\tilde{\theta}\right)=Pr_{0}\left(\theta,\tilde{\theta}\right)\left[\int_{1}^{\infty}pr_{0}\left(t|\theta,\tilde{\theta}\right)dt+\int_{0}^{1}tpr_{0}\left(t|\theta,\tilde{\theta}\right)dt\right].
\end{equation}
Using the normalization integral for $pr_{0}\left(t|\theta,\tilde{\theta}\right)$
we can write
\begin{equation}
P_{e}\left(\theta,\tilde{\theta}\right)=Pr_{0}\left(\theta,\tilde{\theta}\right)\left[1+\int_{0}^{1}\left(t-1\right)pr_{0}\left(t|\theta,\tilde{\theta}\right)dt\right].
\end{equation}
since the integral in this expression is an expectation, it can be
written in terms of an expectation over the data vector $\mathbf{g}$
and we have
\begin{equation}
P_{e}\left(\theta,\tilde{\theta}\right)=Pr_{0}\left(\theta,\tilde{\theta}\right)\left[1-\int_{D}\left[1-t\left(\mathbf{g}|\theta,\tilde{\theta}\right)\right]step\left[1-t\left(\mathbf{g}|\theta,\tilde{\theta}\right)\right]pr\left(\mathbf{g}|\theta\right)d^{M}g\right].
\end{equation}
This is the form for the MPE of the Ziv-Zakai classification task
that we will find most useful.

Now we suppose that $\triangle\theta>0$. The case with negative $\triangle\theta$
will be similar. We will assume that the second derivative of $t\left(\mathbf{g}|\theta,\tilde{\theta}\right)$
with respect to $\tilde{\theta}$ is continuous and note that $t\left(\mathbf{g}|\theta,\theta\right)=1$.
Using Taylor's Theorem with Remainder we find that, for $\triangle\theta$
small enough we can write
\begin{equation}
step\left[1-t\left(\mathbf{g}|\theta,\theta+\triangle\theta\right)\right]=step\left[-t'\left(\mathbf{g}|\theta\right)\triangle\theta\right]=step\left[-t'\left(\mathbf{g}|\theta\right)\right].
\end{equation}
Therefore we have the expansion
\begin{equation}
P_{e}\left(\theta,\theta+\triangle\theta\right)=\left[\frac{1}{2}-\frac{pr'\left(\theta\right)}{4pr\left(\theta\right)}\triangle\theta\right]\left[1+\int_{\mathcal{A}}\left[t'\left(\mathbf{g}|\theta\right)\triangle\theta\right]\left[pr\left(\mathbf{g}|\theta\right)+pr'\left(\mathbf{g}|\theta\right)\triangle\theta\right]d^{M}g\right]+\ldots,
\end{equation}
 where $\mathcal{D}$ is the data domain and $\mathcal{A}$ is the
subset of data vectors in $\mathcal{D}$ satisfying $t'\left(\mathbf{g}|\theta\right)<0$.
Keeping only the zero and first order terms we have
\begin{equation}
P_{e}\left(\theta,\theta+\triangle\theta\right)-\frac{1}{2}=\left[-\frac{pr'\left(\theta\right)}{4pr\left(\theta\right)}+\frac{1}{2}\int_{\mathcal{A}}t'\left(\mathbf{g}|\theta\right)pr\left(\mathbf{g}|\theta\right)d^{M}g\right]\triangle\theta+\ldots
\end{equation}
We may write this with an integral over the whole data domain as
\begin{equation}
P_{e}\left(\theta,\theta+\triangle\theta\right)-\frac{1}{2}=\left[-\frac{pr'\left(\theta\right)}{4pr\left(\theta\right)}+\frac{1}{2}\int_{\mathcal{D}}\left\{ 1-step\left[t'\left(\mathbf{g}|\theta\right)\right]\right\} t'\left(\mathbf{g}|\theta\right)pr\left(\mathbf{g}|\theta\right)d^{M}g\right]\triangle\theta+\ldots
\end{equation}
Using the definition $t'\left(\mathbf{g}|\theta\right)$ of this last
expression simplifies to 
\begin{equation}
P_{e}\left(\theta,\theta+\triangle\theta\right)-\frac{1}{2}=-\left[\frac{pr'\left(\theta\right)}{4pr\left(\theta\right)}+\frac{1}{2}\int_{\mathcal{D}}step\left[t'\left(\mathbf{g}|\theta\right)\right]t'\left(\mathbf{g}|\theta\right)pr\left(\mathbf{g}|\theta\right)d^{M}g\right]\triangle\theta+\ldots
\end{equation}
This will be the MPE expression we will use going forward.

Now we need to take an expectation over $\theta$ using the prior
PDF $pr\left(\theta\right)$. The result is
\begin{equation}
\left\langle P_{e}\left(\theta,\theta+\triangle\theta\right)-\frac{1}{2}\right\rangle _{\theta}=-\frac{1}{2}\left\langle \int_{\mathcal{D}}step\left[t'\left(\mathbf{g}|\theta\right)\right]t'\left(\mathbf{g}|\theta\right)pr\left(\mathbf{g}|\theta\right)d^{M}g\right\rangle _{\theta}\triangle\theta+\ldots
\end{equation}
In terms of expectations we can now write
\begin{equation}
\left\langle 1-2P_{e}\left(\theta,\theta+\triangle\theta\right)\right\rangle _{\theta}=\left\langle \left\langle t'step\left(t'\right)\right\rangle _{\mathbf{g}|\theta}\right\rangle _{\theta}\triangle\theta+\ldots
\end{equation}
By considering $\triangle\theta<0$ case we can summarize the results
in terms of one-side derivatives as
\begin{equation}
\left\langle \left.\frac{d}{d\tilde{\theta}}\left[1-2P_{e}\left(\theta,\tilde{\theta}\right)\right]\right|_{\tilde{\theta}=\theta\pm}\right\rangle _{\theta}=\pm\left\langle \left\langle t'step\left(t'\right)\right\rangle _{\mathbf{g}|\theta}\right\rangle _{\theta}.
\end{equation}
Since the mean of $t'$ is zero we have the final result
\begin{equation}
\left\langle \left.\frac{d}{d\tilde{\theta}}\left[1-2P_{e}\left(\theta,\tilde{\theta}\right)\right]\right|_{\tilde{\theta}=\theta\pm}\right\rangle _{\theta}=\pm\frac{1}{2}\left\langle \left\langle \left|t'\right|\right\rangle _{\mathbf{g}|\theta}\right\rangle _{\theta}.
\end{equation}
This is similar to the result that we discussed in Section 3 relating
FI to the ideal observer AUC for this classification task. For fixed
$\theta$ the function $1-2P_{e}\left(\theta,\tilde{\theta}\right)$
reaches a minimum value of zero at $\tilde{\theta}=\theta$ but it
is not differentiable there. The slope as we move away from this singularity
is determined by the mean value of the random variable $\left|t'\right|$.
Now, using the Schwarz inequality we can bring in the Bayesian FI
as follows
\begin{equation}
\left|\left\langle \left.\frac{d}{d\tilde{\theta}}\left[1-2P_{e}\left(\theta,\tilde{\theta}\right)\right]\right|_{\tilde{\theta}=\theta\pm}\right\rangle _{\theta}\right|\leq\frac{1}{2}\sqrt{F}.
\end{equation}
We will see in the examples below that it is possible to have equality
in this relation. In terms of the MPE for the Ziv-Zakai task directly
we may write the one-sided derivatives as
\begin{equation}
\left.\frac{d}{d\tilde{\theta}}\left\langle P_{e}\left(\theta,\tilde{\theta}\right)\right\rangle _{\theta}\right|_{\tilde{\theta}=\theta\pm}=\mp\frac{1}{4}\left\langle \left\langle \left|t'\right|\right\rangle _{\mathbf{g}|\theta}\right\rangle _{\theta}.
\end{equation}
The Bayesian FI inequality then has the form
\begin{equation}
\left|\left.\frac{d}{d\tilde{\theta}}\left\langle P_{e}\left(\theta,\tilde{\theta}\right)\right\rangle _{\theta}\right|_{\tilde{\theta}=\theta\pm}\right|\leq\frac{1}{4}\sqrt{F}.
\end{equation}
This nequality completes the circle relating MPE for the Ziv-Zakai
calssification task, EMSE for a parameter estimator, and the Bayesian
FI. It also provides a new interpretation of the Bayesian FI in terms
of the average MPE for the task of detecting a small change in a parameter. 

As a final note in this section we can rearrange the expectations
in the mean of $\left|t'\right|$ and find that 
\begin{equation}
\left\langle \left\langle \left|t'\right|\right\rangle _{\mathbf{g}|\theta}\right\rangle _{\theta}=\left\langle \left\langle \left|t'\right|\right\rangle _{\theta|\mathbf{g}}\right\rangle _{\mathbf{g}}=\left\langle \int_{-\infty}^{\infty}\left|pr'\left(\theta|\mathbf{g}\right)\right|d\theta\right\rangle _{\mathbf{g}}.
\end{equation}
This last expectation is the average value of the total variation
(TV) of the posterior PDF $pr\left(\theta|\mathbf{g}\right)$. It
is this quantity which governs the behavior of $\left\langle P_{e}\left(\theta,\tilde{\theta}\right)\right\rangle _{\theta}$
when $\tilde{\theta}$ is close to $\theta$. From standard results
about the total variation we then have
\begin{equation}
\left.\frac{d}{d\tilde{\theta}}\left\langle P_{e}\left(\theta,\tilde{\theta}\right)\right\rangle _{\theta}\right|_{\tilde{\theta}=\theta\pm}\leq-\frac{1}{4}\left\langle \frac{1}{N}\sum_{n=1}^{N}\left|pr\left(\theta_{n}|\mathbf{g}\right)-pr\left(\theta_{n-1}|\mathbf{g}\right)\right|\right\rangle _{\mathbf{g}}
\end{equation}
and
\begin{equation}
\left.\frac{d}{d\tilde{\theta}}\left\langle P_{e}\left(\theta,\tilde{\theta}\right)\right\rangle _{\theta}\right|_{\tilde{\theta}=\theta-}\geq\frac{1}{4}\left\langle \frac{1}{N}\sum_{n=1}^{N}\left|pr\left(\theta_{n}|\mathbf{g}\right)-pr\left(\theta_{n-1}|\mathbf{g}\right)\right|\right\rangle _{\mathbf{g}}.
\end{equation}
For large values of $N$ the sums can give us a good approximation
to the TV of the posterior PDF when it cannot be computed analytically.
The TV of the posterior PDF can be thought of as an average figure
of merit for the task of detecting a small change in a parameter that
is also related, via the Bayesian FI and the van Trees inequality,
to the EMSE of an estimator for the same parameter. 

\section{Vector Version for Bayesian FIM and MPE}

The results of the previous sectipon can also be extended to $p$-dimensional
vector parameters. The Bayesian FIM is the$p\times p$ matrix

\begin{equation}
\mathbf{F}=\left\langle \left\langle \left[\nabla_{\boldsymbol{\theta}}\ln pr\left(\boldsymbol{\theta}|\mathbf{g}\right)\right]\left[\nabla_{\boldsymbol{\theta}}\ln pr\left(\boldsymbol{\theta}|\mathbf{g}\right)\right]^{\dagger}\right\rangle _{\mathbf{g}|\boldsymbol{\theta}}\right\rangle _{\boldsymbol{\theta}}.
\end{equation}
If $\mathbf{u}$is a unit vector in the parameter space, then we can
show that the one sided directional derivatives satisfy
\begin{equation}
\frac{d}{dt}\left.\left\langle P_{e}\left(\boldsymbol{\theta},\boldsymbol{\theta}+t\mathbf{u}\right)\right\rangle _{\theta}\right|_{t=0+}\geq-\frac{1}{4}\sqrt{\mathbf{u}^{\dagger}\mathbf{Fu}}
\end{equation}
and
\begin{equation}
\frac{d}{dt}\left.\left\langle P_{e}\left(\boldsymbol{\theta},\boldsymbol{\theta}+t\mathbf{u}\right)\right\rangle _{\theta}\right|_{t=0-}\leq\frac{1}{4}\sqrt{\mathbf{u}^{\dagger}\mathbf{Fu}}.
\end{equation}
Therefore the components of the Bayesian FIM are related to the change
in the average MPE $\left\langle P_{e}\left(\boldsymbol{\theta},\boldsymbol{\theta}'\right)\right\rangle _{\boldsymbol{\theta}}$
as $\boldsymbol{\theta}'$ moves away from $\boldsymbol{\theta}$.
There are vector versions of the van Trees inequality and the Ziv-Zakai
inequality, so in the vectro case also the EMSE, the Bayesian FIM
and the MPE for the task of detecting a change in the parameter vector
are all related.

\section{Unimodal Posterior PDF}

MNow we return to the scalar parameter case and suppose that the posterior
PDF $pr\left(\theta|\mathbf{g}\right)$ is supported on the (possibly
infinite) interval between $a$ and $b$. If this PDF is unimodal
with mode $x$, then the TV of the posterior can be written as

\begin{equation}
\int_{a}^{b}\left|pr'\left(\theta|\mathbf{g}\right)\right|d\theta=\int_{a}^{x}pr'\left(\theta|\mathbf{g}\right)d\theta-\int_{x}^{b}pr'\left(\theta|\mathbf{g}\right)d\theta.
\end{equation}
This expression is the same as
\begin{equation}
\int_{a}^{b}\left|pr'\left(\theta|\mathbf{g}\right)\right|d\theta=2\int_{a}^{x}pr'\left(\theta|\mathbf{g}\right)d\theta-\int_{a}^{b}pr'\left(\theta|\mathbf{g}\right)d\theta,
\end{equation}
and, since the last term is zero ,we have
\begin{equation}
\int_{a}^{b}\left|pr'\left(\theta|\mathbf{g}\right)\right|d\theta=2\int_{a}^{x}pr'\left(\theta|\mathbf{g}\right)d\theta=2pr\left(x|\mathbf{g}\right).
\end{equation}
Thus the TV for the posterior PDF is easy to calculate in this case
if we have an analytic expression for this PDF. 

\section{Two Examples of Unimodal Posterior PDFs}

We consider two examples of possible unimodal posteriror PDFs. The
first is a normal distribution:
\begin{equation}
pr\left(\theta|m,\sigma\right)=\frac{1}{\sqrt{2\pi\sigma^{2}}}\exp\left[-\frac{1}{2\sigma^{2}}\left(\theta-m\right)^{2}\right].
\end{equation}
In this case we have for the posterior TV
\begin{equation}
\int_{-\infty}^{\infty}\left|pr'\left(\theta|m,\sigma\right)\right|d\theta=\frac{1}{\sigma}\sqrt{\frac{2}{\pi}}.
\end{equation}
The corresponding vaslue for the Bsyesian FI is given by
\begin{equation}
\left\langle \left[\frac{pr'\left(\theta|m,\sigma\right)}{pr\left(\theta|m,\sigma\right)}\right]^{2}\right\rangle _{\theta|m,\sigma}=\frac{1}{\sigma^{2}}.
\end{equation}
Thus the ratrio of the posterior TV to the square root of the Bayesian
FI is $\sqrt{2/\pi}$. This number is less than unity as expected.

The second example of a possible posterior PDF exponential distribution:
\begin{equation}
pr\left(\theta|\beta\right)=\beta\exp\left(-\beta\theta\right).
\end{equation}
In this case the posterior TV is given by
\begin{equation}
\int_{0}^{\infty}\left|pr'\left(\theta|\beta\right)\right|d\theta=\beta,
\end{equation}
while the Bayesian FI is
\begin{equation}
\left\langle \left[\frac{pr'\left(\theta|\beta\right)}{pr\left(\theta|\beta\right)}\right]^{2}\right\rangle _{\theta|\alpha,\beta}=\beta^{2}.
\end{equation}
The ratio of the posterior TV to the square root of the Bayesian FI
for this example is unity. Since there are many cases where an exponential
distribution is a posterior PDF this example shows that it is possible
to have equality in the relation derived in Section 7 between the
square root of the Bayesian FI and the one-sided derivatives of the
MPE for the task of detecting a small change in a parameter.

\section{Example: Multivariate Gaussian and Gaussian}

Now we consider an example with a scalar parameter and Gaussian statistics
for the conditional PDF $pr\left(\mathbf{g}|\theta\right)$ and the
prior PDF $pr\left(\theta\right)$. Specifically we fix a unit vector
$\mathbf{s}$in data space and assume that the conditional PDF is
given by

\begin{equation}
pr\left(\mathbf{g}|\theta\right)=\frac{1}{\sqrt{2\pi\det\mathbf{K}}}\exp\left[-\frac{1}{2}\left(\mathbf{g}-\theta\mathbf{s}\right)^{\dagger}\mathbf{K}^{-1}\left(\mathbf{g}-\theta\mathbf{s}\right)\right].
\end{equation}
We may decompose the data vector as $\mathbf{g}=\mathbf{g}_{\bot}+\mathbf{g}_{\Vert}$
with $\mathbf{g}_{\bot}^{\dagger}\mathbf{K}^{-1}\mathbf{s}=0$ and
$\mathbf{g}_{\Vert}=g_{\Vert}\mathbf{s}$. Then the conditional PDF
may be written as
\begin{equation}
pr\left(\mathbf{g}|\theta\right)=\frac{1}{\sqrt{2\pi\det\mathbf{K}}}\exp\left[-\frac{1}{2}\left(g_{\Vert}-\theta\right)^{\dagger}\mathbf{s}^{\dagger}\mathbf{K}^{-1}\mathbf{s}\left(g_{\Vert}-\theta\right)\right]\exp\left[-\frac{1}{2}\mathbf{g}_{\bot}^{\dagger}\mathbf{K}^{-1}\mathbf{g}_{\bot}\right].
\end{equation}
We will assume a Gaussian prior PDF:
\begin{equation}
pr\left(\theta\right)=\frac{1}{\sqrt{2\pi\sigma^{2}}}\exp\left[-\frac{1}{2\sigma^{2}}\left(\theta-\mu\right)^{2}\right].
\end{equation}
The posterior PDF is also a Gaussian and all that we need is its variance
$\sigma_{p}^{2}$ , which is given by
\begin{equation}
\frac{1}{\sigma_{p}^{2}}=\mathbf{s}^{\dagger}\mathbf{K}^{-1}\mathbf{s}+\frac{1}{\sigma^{2}}.
\end{equation}
Now for the posterior TV we have
\begin{equation}
\int_{-\infty}^{\infty}\left|pr'\left(\theta|\mathbf{g}\right)\right|d\theta=\sqrt{\frac{2}{\pi\sigma_{p}^{2}}}=\sqrt{\frac{2}{\pi}\left(\mathbf{s}^{\dagger}\mathbf{K}^{-1}\mathbf{s}+\frac{1}{\sigma^{2}}\right).}Since
\end{equation}
this quantity does not depend on $\mathbf{g}$ we have or the average
posterior TV
\begin{equation}
\left\langle \left\langle \left|t'\left(\mathbf{g}|\theta\right)\right|\right\rangle _{\mathbf{g}|\theta}\right\rangle _{\theta}=\sqrt{\frac{2}{\pi}\left(\mathbf{s}^{\dagger}\mathbf{K}^{-1}\mathbf{s}+\frac{1}{\sigma^{2}}\right)}.
\end{equation}
The Bayesian FI is given by
\begin{equation}
F=\mathbf{s}^{\dagger}\mathbf{K}^{-1}\mathbf{s}+\frac{1}{\sigma^{2}}.
\end{equation}
Therefore the ratio of the average posterior TV and the square root
of the Bayesian FI is $\sqrt{2/\pi}$. 

\section{Example: Multivariate Gausssians}

In this example we consider a conditional PDF of the following form:

\begin{equation}
pr\left(\mathbf{g}|\boldsymbol{\theta}\right)=\frac{1}{\sqrt{2\pi\det\mathbf{K}_{n}}}\exp\left[-\frac{1}{2}\left(\mathbf{g}-\mathbf{H}\boldsymbol{\theta}\right)^{\dagger}\mathbf{K}_{n}^{-1}\left(\mathbf{g}-\mathbf{H}\boldsymbol{\theta}\right)\right].
\end{equation}
We may think of the $M\times N$ matrix $\mathbf{H}$ as representing
an imaging system acting on the $N$- dimensional object vector $\boldsymbol{\theta}$
and generating the $M$-dimensional noisy data vector $\mathbf{g}$,
where the noise is described by correlated gaussian statistics. We
will assume that $M<N$ and that the matrix $\mathbf{H}$ is full
rank. If $\mathbf{H}^{+}$ is the pseudoinverse of $\mathbf{H}$,
then these assumptions imply that $\mathbf{H}\mathbf{H}^{+}=\mathbf{I}$.
We may therefore write the conditional PDF as
\begin{equation}
pr\left(\mathbf{g}|\boldsymbol{\theta}\right)=\frac{1}{\sqrt{2\pi\det\mathbf{K}_{n}}}\exp\left[-\frac{1}{2}\left(\boldsymbol{\theta}-\mathbf{H}^{+}\mathbf{g}\right)^{\dagger}\mathbf{H}^{\dagger}\mathbf{K}_{n}^{-1}\mathbf{H}\left(\boldsymbol{\theta}-\mathbf{H}^{+}\mathbf{g}\right)\right].
\end{equation}
We assume the pior PDF is also Gaussian and given by
\begin{equation}
pr\left(\boldsymbol{\theta}\right)=\frac{1}{\sqrt{2\pi\det\mathbf{K}}}\exp\left[-\frac{1}{2}\left(\boldsymbol{\theta}-\boldsymbol{\mu}\right)^{\dagger}\mathbf{K}_{\theta}^{-1}\left(\boldsymbol{\theta}-\boldsymbol{\mu}\right)\right].
\end{equation}
It is now easy to see that the posterior PDF is also Gaussian with
a covariance matrix given by $\mathbf{K}_{p}^{-1}=\mathbf{K}_{\theta}^{-1}+\left(\mathbf{H}^{\dagger}\mathbf{K}_{n}^{-1}\mathbf{H}\right)$.
If we define a directional derivative for the unit vector $\mathbf{u}$in
parameter space by
\begin{equation}
D_{\mathbf{u}}pr\left(\boldsymbol{\theta}|\mathbf{g}\right)=\left.\frac{d}{dt}pr\left(\boldsymbol{\theta}+t\mathbf{u}|\mathbf{g}\right)\right|_{t=0},
\end{equation}
then the vector analogue of the posterior TV for the case of a scalar
parameter is given by
\begin{equation}
\int_{\mathbb{R}^{N}}\left|D_{\mathbf{u}}pr\left(\boldsymbol{\theta}|\mathbf{g}\right)\right|d^{N}\theta=\int_{\mathbb{R}^{N}}\left|\mathbf{u}^{\dagger}\mathbf{K}_{p}^{-1}\left(\boldsymbol{\theta}-\boldsymbol{\mu}_{p}\right)\right|pr\left(\boldsymbol{\theta}|\mathbf{g}\right)d^{N}\theta.
\end{equation}
This integral can be computed and reduces to
\begin{equation}
\int_{\mathbb{R}^{N}}\left|D_{\mathbf{u}}pr\left(\boldsymbol{\theta}|\mathbf{g}\right)\right|d^{N}\theta=\sqrt{\frac{2}{\pi}\mathbf{u}^{\dagger}\mathbf{K}_{p}^{-1}\mathbf{u}}.
\end{equation}
As in the scalar case, this number is the magnitude of the one-sided
directional derivative of $P_{e}\left(\boldsymbol{\theta},\boldsymbol{\theta}'\right)$
when $\boldsymbol{\theta}'$ is moving away from $\boldsymbol{\theta}$
in the direction . The square root of the $\mathbf{u}$ component
of the Bayesian FIM is given by
\begin{equation}
\sqrt{\mathbf{u}^{\dagger}\mathbf{F}\mathbf{u}}=\sqrt{\mathbf{u}^{\dagger}\mathbf{K}_{p}^{-1}\mathbf{u}}.
\end{equation}
Ghe ratio of these two quantities is once again $\sqrt{2/\pi}$. This
reflects the fact that the multivariate version of the posterior TV
in the direction $\mathbf{u}$will always be less than or equal to
$\sqrt{\mathbf{u}^{\dagger}\mathbf{F}\mathbf{u}}$, a fact which can
be proved using the same methods used above for the scalar case. 

\section{Conclusion}

The relation between FI and the ideal-observer AUC described above
relates the FI to the ability of the ideal observer to detect a small
change in a scalar parameter that is affecting the statistics of the
data vector. The relation between the FIM and the ideal observer AUC
is similar except that we are trying to detect a change in a vector
parameter. In both cases the AUC is approximately proportional to
the square root of the relevant component of the FIM for small changes
in the parameters. In this work we wanted to extend these results
to the Bayesian FI for scalar parameters and the Bayesian FIM for
vector parameters. . The ideal-observer AUC and the FIM do not depend
on the prior probabilities for the Signal-Present and Signal Absent
hypotheses. The new element in the Bayesian approach is a prior on
the parameters governing the statistics of the data, which can be
used to define these prior probabilites.

This extension is based on a task introduced in the Ziv-Zakai inequality
where the ideal observer is trying to detect a change in a parameter
and the prior probabilities for the two hypotheses are determined
by the prior PDF on the parameter. In this case the AUC for the ideal-observer
is no longer relevant and it is the probability of error for the ideal
observer, the MPE, that becomes the detection figure of merit. The
Ziv-Zakai inequality relates the MPE to the EMSE for any estimator
of the parameter. In this work we related this MPE for small deviations
in the parameter to the Bayesian FIM via an inequality. An example
shows that this inequality will be equality for certian posterior
PDFs, so in this sense the inequality is sharp. An intermediate quantity
in the derivation of this result is the posterior TV, which is related
to the small-deviation MPE as the first term in a Taylor series expansion.
This relation is similar to the relation between ideal-observer AUC
and FI discussed above. 

The results discussed in this work further elucidate the connections
between estimating a parameter and detcting a change in that parameter.
An imaging system optimized for one of these tasks will probably be
optimized for the other. In particular, if we are using FI or the
Bayesian FI for optimization on an estimation task, then we are also
optimizing for the task of detecting a small change in the parameter
of interest. The results in this paper and others {[}11,12{]} also
show that the well known measures of information, ideal-observer AUC,
MPE, FI, Bayesian FI and Shannon Information, are all related to each
other in ways that are not always obvious. We may also now add the
posterior TV to this list as a measure of information related to both
detection and estimation tasks. 

\section{Appendix}

Here we show the steps that lead to our alternate form for the Ziv-Zakai
inequality. By making a simple change of variables we may convert
the usual version, given above, to the inequality

\begin{equation}
EMSE\geq\frac{1}{2}\int_{-\infty}^{\infty}\int_{\theta}^{\infty}\left[pr\left(\theta\right)+pr\left(\tilde{\theta}\right)\right]P_{e}\left(\theta,\tilde{\theta}\right)\left(\tilde{\theta}-\theta\right)d\tilde{\theta}d\theta.
\end{equation}
Due to the limits of integration on the inner integral we may write
this inequality as
\begin{equation}
EMSE\geq\frac{1}{2}\int_{-\infty}^{\infty}\int_{\theta}^{\infty}\left[pr\left(\theta\right)+pr\left(\tilde{\theta}\right)\right]P_{e}\left(\theta,\tilde{\theta}\right)\left|\tilde{\theta}-\theta\right|d\tilde{\theta}d\theta.
\end{equation}
Now we interchange the order of integration to write
\begin{equation}
EMSE\geq\frac{1}{2}\int_{-\infty}^{\infty}\int_{-\infty}^{\tilde{\theta}}\left[pr\left(\theta\right)+pr\left(\tilde{\theta}\right)\right]P_{e}\left(\theta,\tilde{\theta}\right)\left|\tilde{\theta}-\theta\right|d\theta d\tilde{\theta}.
\end{equation}
We can use the fact that $P_{e}\left(\theta,\tilde{\theta}\right)=P_{e}\left(\tilde{\theta},\theta\right)$
and rename the integration variables to get
\begin{equation}
EMSE\geq\frac{1}{2}\int_{-\infty}^{\infty}\int_{-\infty}^{\theta}\left[pr\left(\theta\right)+pr\left(\tilde{\theta}\right)\right]P_{e}\left(\theta,\tilde{\theta}\right)\left|\tilde{\theta}-\theta\right|d\tilde{\theta}d\theta.
\end{equation}
Combining the second and fourth inequalities in this Appendix now
gives
\begin{equation}
EMSE\geq\frac{1}{4}\int_{-\infty}^{\infty}\int_{-\infty}^{\infty}\left[pr\left(\theta\right)+pr\left(\tilde{\theta}\right)\right]P_{e}\left(\theta,\tilde{\theta}\right)\left|\tilde{\theta}-\theta\right|d\tilde{\theta}d\theta.
\end{equation}
Finally splitting this into two integrals, and using the symmetry
of the MPE again to realize that the two integrals are the same, gives
us the end result
\begin{equation}
EMSE\geq\frac{1}{2}\int_{-\infty}^{\infty}\int_{-\infty}^{\infty}pr\left(\theta\right)P_{e}\left(\theta,\tilde{\theta}\right)\left|\tilde{\theta}-\theta\right|d\tilde{\theta}d\theta.
\end{equation}
This last expression can be written as an expectation as in the main
text above. 

\section{References}

\noun{1 .J. Shao}, \emph{Mathematical Statistics,} Springer, New York
(1999).

2. H. H. Barrett, K. J. Myers, \emph{Foundations of Image Science},
John Wiley \& Sons, Hoboken, NJ (2004).

3. H. Barrett, C. Abbey, E. Clarkson, \textquotedbl{}Objective assessment
of image quality. III. ROC metrics, ideal observers, and likelihood-generating
functions,\textquotedbl{} J. Opt. Soc. Am. A 15, 1520-1535 (1998). 

4. E. Clarkson , F. Shen, ``Fisher information and surrogate figures
of merit for the task-based assessment of image qualit\emph{y},''
JOSA A 27, 2313-2326 (2010) . 

5. F. Shen, E. Clarkson, ``Using Fisher information to approximate
ideal observer performance on detection tasks for lumpy-background
images,'' JOSA A 23, 2406-2414 (2006). 

6. E. Clarkson, \textquotedbl{}Asymptotic ideal observers and surrogate
figures of merit for signal detection with list-mode data,\textquotedbl{}
J. Opt. Soc. Am. A 29, 2204-2216 (2012). 

7. J. Ziv, M. Zakai, \textquotedbl{}Some Lower Bounds on Signal Parameter,\textquotedbl{}
IEEE Trans. on Information Theory 15, 386-391 (1969).

8. K. Bell, Y. Steinberg, Y. Ephraim, and H. Van Trees, ``Extended
Ziv-Zakai lower bound for vector parameter estimation,'' IEEE Trans.
on Information Theory 43, 624\textendash{} 637 (1997).

9. H. L. van Trees, \emph{Detection, Estimation and Modulation Theory},
Part 1, New York, Wiley, (1968).

10. R.D. Gill, B.Y. Levit, ``Applications of the van Trees inequality:
a Bayesian Cramér\textendash Rao bound, `` Bernoulli 1, 59\textendash 79
(1995).

11. E. Clarkson, J. Cushing, \textquotedbl{}Shannon information and
ROC analysis in imaging,\textquotedbl{} J. Opt. Soc. Am. A 32, 1288-1301
(2015). 

12. E. Clarkson, J. Cushing, \textquotedbl{}Shannon information and
receiver operating characteristic analysis for multiclass classification
in imaging,\textquotedbl{} J. Opt. Soc. Am. A 33, 930-937 (2016). 
\end{document}